\documentclass[aps,pra,twocolumn,superscriptaddress,longbibliography]{revtex4-1}
\usepackage[T1]{fontenc}
\usepackage{graphicx}
\usepackage{amsmath}
\usepackage{amsfonts}
\usepackage{multirow}
\usepackage{makecell}
\usepackage[table,xcdraw]{xcolor}
\usepackage{times}
\usepackage[hidelinks]{hyperref}
\usepackage{chngcntr}
\hypersetup{
    colorlinks,
    linkcolor={blue!80!black},
    citecolor={blue!80!black},
    urlcolor={blue!80!black}
}
\newcommand{\ket}[1]{\mbox{$\left|#1\right\rangle$}}
\newcommand{\bra}[1]{\mbox{$\left\langle#1\right|$}}

\newcommand{\outerp}[2]{\ket{#1}\bra{#2}}
\DeclareMathOperator{\tr}{\text{Tr}}

\newcommand{\isk}[2]{{ #2} }
\newcommand{\bnd}[2]{{ #2} }

\newcolumntype{?}{!{\vrule width 2pt}}

\begin{document}

\title{Disorder-free localization in quantum walks}

\author{B. Danac\i}
\email{danacib@itu.edu.tr}
\affiliation{Department of Physics, Faculty of Science and Letters, Istanbul Technical University, 34469 Maslak, \.{I}stanbul, Turkey}

\author{\.{I}. Yal\c{c}{\i}nkaya}
\affiliation{Department of Physics, Faculty of Nuclear Sciences and Physical Engineering, Czech Technical University in Prague, B\v{r}ehov\'a 7, 115 19 Praha 1-Star\'e M\v{e}sto, Czech Republic}

\author{B. \c{C}akmak}
\affiliation{College of Engineering and Natural Sciences, Bah\c{c}e\c{s}ehir University, Be\c{s}ikta\c{s}, \.{I}stanbul 34353, Turkey}

\author{G. Karpat}
\affiliation{Faculty of Arts and Sciences, Department of Physics, \.{I}zmir University of Economics, \.{I}zmir, 35330, Turkey}

\author{S. P. Kelly}
\affiliation{Theoretical Division, Los Alamos National Laboratory, Los Alamos, New Mexico 87545, USA}
\affiliation{Department of Physics and Astronomy, University of California Riverside, Riverside, California 92521, USA}

\author{A. L. Suba{\c s}{\i}}
\affiliation{Department of Physics, Faculty of Science and Letters, Istanbul Technical University, 34469 Maslak, \.{I}stanbul, Turkey}
\affiliation{Center for Nonlinear Studies, Los Alamos National Laboratory, Los Alamos, New Mexico 87545, USA}

\date{\today}

\begin{abstract}
The phenomenon of localization usually happens due to the existence of disorder in a medium. Nevertheless, certain quantum systems allow dynamical localization solely due to the nature of internal interactions. We study a discrete time quantum walker which exhibits disorder free localization. The quantum walker moves on a one-dimensional lattice and interacts with on-site spins by coherently rotating them around a given axis at each step. Since the spins do not have dynamics of their own, the system poses the local spin components along the rotation axis as an extensive number of conserved moments. When the interaction is weak, the spread of the walker shows subdiffusive behaviour having downscaled ballistic tails in the evolving probability distribution at intermediate time scales. However, as the interaction gets stronger the walker gets exponentially localized in the complete absence of disorder in both lattice and initial state. Using a matrix-product-state ansatz, we investigate the relaxation and entanglement dynamics of the on-site spins due to their coupling with the quantum walker. Surprisingly, we find that even in the delocalized regime, entanglement growth and relaxation occur slowly unlike marjority of the other models displaying a localization transition.
\end{abstract}


\maketitle

\section{Introduction}

By yielding insights into fundamental questions on thermalization of closed quantum systems, disorder and localization have earned themselves a central place in quantum physics. The corner stone of these topics is the phenomenon of Anderson localization~\cite{anderson1958}, which in one-dimension produces localization of all single particle eigenstates for arbitrarily weak disorder. Due to these localized eigenstates, particles can not move through the system and thermalization is prevented. The absence of transport and thermalization has been found to be robust to interactions~\cite{a1tman2015,nandkishore2015,abanin2017,alet2018,abanin2019}. This many-body localization (MBL), which occurs at finite temperature in the presence of sufficiently strong disorder and interactions, is closely related to the subject of ergodicity-breaking due to the impossibility of thermalization for MBL systems. In fact, the presence of a MBL phase serves as an example of the violation of the eigenstate thermalization hypothesis~\cite{deutsch2018}. Disentangled quantum liquids provide another example of non-thermalizing localization where a light species can localize with low entropy due to the coupling to a system of heavy particles~\cite{Grover_2014, smith2017}. On the other hand, the manifestation of disorder-free localization (DFL) in quantum systems has been demonstrated more recently, where the system generates its own effective disorder dynamically in the absence of any external randomness~\cite{smith2017disorder}. Another example of DFL has been observed in a translationally invariant Ising-Kondo Lattice model, where Anderson localization occurs due to the conserved moments~\cite{yang2020hidden}. 

Having been put forward almost three decades ago as a quantum counterpart of random walks~\cite{aharonov1993}, quantum walks proved themselves so far to be a versatile model for quantum computation, not only due to their role in the development of new quantum algorithms~\cite{ambainis2003} but also for providing a concrete framework for universality~\cite{lovett2010}. Despite their superiority over random walks in spreading rates~\cite{venegas-andraca2012,kempe2003} leading to faster computational algorithms~\cite{ambainis2003}, they have also attracted considerable attention in terms of their non-diffusibility features in the presence of disorder such as quantum-to-classical transition~\cite{ahlbrecht2011asymptotic}, dynamical localization~\cite{ahlbrecht2011disordered,hamza2009dynamical}, and the emergence of bound states~\cite{wojcik2012trap,danaci2019non}. While rolling back to the classical behavior is attributed to wiping out coherence in the system due to dynamical disorder, it is the static disorder that leads to dynamical localization in quantum walks. The former disorder scheme manifests itself as a time-dependent evolution operator, whereas the latter resembles a random potential for the quantum walker described by a unitary operator~\cite{torma2002,mackay2002,brun2003coins,brun2003walks,annabestani2010,romanelli2005decoherence,singh2020}. Since these disorder models break either the temporal or the spatial symmetries, any emergent non-diffusibility in quantum walks is typically associated with a lack of some symmetry in the dynamics~\cite{chandrashekar2007,yin2008,chandrashekar2011,chandrashekar2013,schreiber2011}. Other examples of non-diffusibility have been observed in quantum walks coupled to spin environments~\cite{prokof2006decoherence,PhysRevLett.116.247202}.

In this article we study DFL in a quantum walk model. In particular, we consider a discrete-time walker coupled to on-site spin-half systems on a one-dimensional lattice such that the presence of the walker on a specific site coherently rotates the corresponding spin. Our examination of the considered system involves two parts. First, we consider the walker and investigate its spreading dynamics as it interacts with the spins. For weak coupling the probability distribution is only partially localized around the origin and still has ballistic tails spreading out linearly in time. As the coupling gets stronger, the walker becomes completely localized with no ballistic tails in the time evolution. This dynamical localization appears without breaking any translational or temporal symmetries.

In this sense, our model provides a simple manifestation of interaction-induced disorder-free localization~\cite{smith2017disorder}. We then turn our attention to the spin system and study its relaxation and entanglement dynamics. Similar to MBL, the local integrals of motion on the spin chain decohere and entangle due to emergent interactions between them. When the walker is localized, the emergent interaction decays exponentially and leads to sublinear growth in entanglement mimicking the effect in MBL systems. Interestingly, for our simulations at intermediate time scales, entanglement growth remains slow also in the seemingly delocalized regime due to the suppression of the ballistic tails and the low coupling.

This manuscript is organized as follows. In Sec.~\ref{sec:model}, we introduce our model and discuss the implications of the extensive number of local symmetries for disorder-free localization. Our main results are presented in Sec.~\ref{sec:results}. We first provide the quasi-energy spectrum as a function of the interaction parameter and study the localization properties of the walker and dynamics of the local spins. Then, we demonstrate the logarithmic growth of entanglement in time that arises due to an effective interaction between the spins induced by the walker. In Sec. \ref{sec:conclusion}, we conclude by making connections to similarly behaving systems and possible extensions. The equilibration of entanglement entropy for small systems, analysis of the entanglement spread, details of the matrix-product-state (MPS) calculations, and the perturbative effects of a symmetry-breaking field are discussed in the appendices.

\section{\label{sec:model} The model}

\subsection{Discrete-time quantum walk\label{sec:walk}}
Analogous to the classical random walk, the discrete-time quantum walk depicts the unitary evolution of a particle on a lattice where the direction of the movement is determined by an internal degree of freedom, namely, the coin \cite{aharonov1993}. At each time step, an operator $\hat{C}$ acting solely on the coin state is followed by a coin-state-conditioned translation operator $\hat{T}$ acting on the total coin-position Hilbert space $\mathcal{H}_c~\otimes~\mathcal{H}_p$. Although $\hat{C}$ can be chosen from $SU(2)$, without loss of generality, one may constrain the coin operator to be a rotation around the $x$-axis~\cite{tregenna2003}, so that 
\begin{equation}
\hat{C} = \exp\left(-i \theta \hat{X}\right) \otimes \hat{I}_N.
\end{equation}
Here, $\hat{X}$ is the Pauli-X operator acting on the coin space and $\hat{I}_N$ is the identity operator acting on the the $N$-dimensional position space. We will further fix the coin angle to be $\theta~=~\pi/4$ throughout the manuscript. The conditional translation operator shifts the position of the walker one site to its right or left depending on the coin state as given by
\begin{equation}
\hat{T} =
\sum_{c,n} 
\outerp{c}{c} \otimes \outerp{n+(-1)^c}{n},
\label{eq:shift}
\end{equation}
where $c\in\{0,1\}$ refer to the eigenstates of the Pauli-Z operator $\hat{Z}$ spanning the coin space and $n \in \mathbb{Z}$ labels the $N$ position states. Thus, the evolution operator describing a single time step in the dynamics is written as $\hat{W}~=~\hat{T}\hat{C}$. After $t$ steps, the final state becomes
\begin{equation}
\ket{\psi_t} = \hat{W}^t \ket{\psi_0} = \sum_{c,n} a_{c,n}(t) \ket{c,n},
\end{equation}
where $\ket{\psi_0}$ is an arbitrary initial state at $t=0$ and $a_{c,n}(t)$ are the probability amplitudes corresponding to coin state $\ket{c}$ and position state $\ket{n}$. Consequently, the probability of finding the quantum walker at position $n$ and step $t$ is calculated by $P_n(t)=\sum_c \left|a_{c,n}(t)\right|^2$. Throughout the manuscript, we consider $a_{c,n\neq 0}(0)=0$ such that $\ket{\psi_0}=\ket{\chi_0}~\otimes~\ket{n=0}$ with the coin state $\ket{\chi_0}~=~2^{-1/2}\left(\ket{0}+\ket{1}\right)$. This initial state is localized at the origin and yields a symmetric probability distribution over the lattice in time with our aforementioned choice of the coin operator~\cite{kempe2003,venegas-andraca2012}.

\subsection{Quantum walk interacting with on-site local spins\label{sec:walkM}}
\begin{figure}[t]
\centering
\includegraphics[scale=1.05]{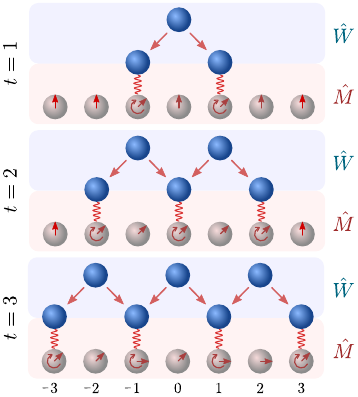}
\caption{\label{fig:theModel}Schematic representation of the first $3$ steps of the quantum walk model employed in the manuscript. The quantum walker (blue-dark balls) interacts with the spin chain (gray-light balls) in each step such that the spin rotates by an angle $\phi$ when the walker is on that site and has no dynamics otherwise. The starting point of the walk is indicated by $0$ and a balanced walk is depicted.}
\end{figure}

In this article, we consider a quantum walker interacting with a chain of $N$ spins which are permanently localized at the lattice sites and not directly coupled with each other. We denote the state of a single spin at site $n$ by $\ket{s_n}$ where $s_n$ will be either $\{0,1\}$ or $\{+,-\}$ referring to the eigenstates of $\hat{Z}$ or $\hat{X}$, respectively. The total Hilbert space is $\mathcal{H}_c~\otimes~\mathcal{H}_p~\otimes~\mathcal{H}_s$, and has size $2\times N\times 2^N$ which scales exponentially in $N$. The spins' subspace $\mathcal{H}_s$ is spanned by $\{\ket{\mathbf{s}}\}\equiv\{\ket{s_1s_2\hdots s_N}\}$ where $\mathbf{s}$ can have $2^N$ possible values corresponding to different spin configurations. We will employ the notation $\mathbf{s}_\textsc{z}$ and $\mathbf{s}_\textsc{x}$ to refer to eigenbasis of $\hat{Z}$ and $\hat{X}$, respectively. The walker interacts with the spins such that its presence on any site $n$ induces a rotation on the corresponding spin state $\ket{s_n}$ by an angle $\phi$ at each step, as shown in Fig. \ref{fig:theModel}. Without loss of generality, we will consider rotations around the $x$-axis throughout the paper. The site dependent interaction operator in the whole Hilbert space of the system can then be written as
\begin{equation}
\hat{M} = \sum_n \hat{I}_2 \otimes \outerp{n}{n} \otimes \exp\left(-i\phi \hat{X}_n \right),
\label{eq:intOpr}
\end{equation}
where $\phi$ is the rotation angle and the operator
\begin{equation}
\hat{X}_n = \hat{I}_2^{\otimes (n-1)} \otimes \hat{X} \otimes \hat{I}_2^{\otimes (N-n)}
\label{eq:locInt}    
\end{equation}
acts on the spin at the $n^\mathrm{th}$ site. Thus, we introduce the single step operator to involve walker-spin interaction as  $\hat{W}_M~=~\hat{T}\hat{C}\hat{M}$ where both $\hat{T}$ and $\hat{C}$ are now naturally extended with $\hat{I}_{2^N}$ for consistency. Like the coin-state dependent translation operator entangling the coin and position degrees of freedom~\cite{carneiro2005,abal2006}, the position-state dependent spin interaction operator entangles the walker and spins as well. Therefore, the total state of the system at time $t$,
\begin{equation}
\ket{\Psi_t} = \sum_{c,n,\mathbf{s}} a_{c,n,\mathbf{s}}(t) \ket{c,n,\mathbf{s}}\, ,
\label{eq:totState}
\end{equation}
is a superposition of both coin, position and spin state degrees of freedom. In general, even after starting with a product state at $t=0$, it is impossible to factorize any degree of freedom completely at later times. At this point, it is worthy to note that the model we consider here is actually simpler in terms of its construction compared to the other similar models since we consider a one-dimensional position space and the spins have no direct action on the walker ~\cite{camilleri2014quantum,verga2018,verga2020}. The relationship we introduce with Eq. \eqref{eq:intOpr} 
is analogous to that of the coin and position degrees of freedom in the conventional quantum walk. 
Here, we apply a rotation to the local spins depending on the walker's position. 
Unlike spin-spin interactions such as spin exchange, this interaction does not depend on the coin degree of freedom.
.

\subsection{Disorder free localization \label{sec:DFL}} 

In our model, the local spins yield $N$ conserved quantities as can be seen by noting that the step operator
\begin{equation}
\hat{W}_M=\sum_{c,n}\outerp{c}{c}e^{-i\theta\hat{X}}\otimes\outerp{n+(-1)^c}{n}\otimes e^{-i\phi\hat{X}_n}
\end{equation}
is invariant under spin flips, i.e., $[\hat{X}_n,\hat{W}_M]=0$ for all $n$, where we extend $\hat{X}_n\to \hat{I}_{2N}\otimes \hat{X}_n$. These extensive number of conserved local quantities act as a hidden disorder and thus the quantum walker evolving with $\hat{W}_M$ localizes dynamically similar to other DFL mechanisms~\cite{smith2017disorder}. This is clearly seen when looking at the simultaneous eigenstates of both $\hat{X}_n$ and $\hat{W}_M$ which are formed within the subspace spanned by $\{\ket{c,n,\mathbf{s}_\textsc{x}}\}$ with $\mathbf{s}_\textsc{x}$ fixed. This basis block diagonalizes $\hat{W}_M$ such that each $2N\times 2N$ block is labeled by one of $2^N$ possible spin configurations $\mathbf{s}_\textsc{x}$, which can be written as
\begin{equation}
\bra{\mathbf{s}'_\textsc{x}}{\hat{W}_M}\ket{\mathbf{s}_\textsc{x}}=
\begin{cases}
\hat{W}_{\mathbf{s}_\textsc{x}}, &\text{if } \mathbf{s}_\textsc{x}=\mathbf{s}'_\textsc{x} \\
0, & \text{otherwise}
\end{cases}.
\end{equation}
The quantum walker can therefore be viewed as a superposition of walkers evolving in a set of disordered landscapes under $\hat{W}_{\mathbf{s}_\textsc{x}}=\hat{T}\hat{C}\hat{D}_{\mathbf{s}_\textsc{x}}$  with the disorder operator $\hat{D}_{\mathbf{s}_\textsc{x}}~\equiv~\sum_n \hat{I}_2\otimes\left|n\right>\left<n\right|e^{i\phi s_n}$, where $s_n$ are the elements of the set $\mathbf{s}_\textsc{x}$~\footnote{Even though the diagonalization problem of $\hat{W}_M$ can be broken down to those of $\hat{W}_{\mathbf{s}_\textsc{x}}$, the number of the small size problems grows exponentially with the system size ($\sim 2^N$) in general. We note that the one-dimensional geometry together with local couplings allows simulations of large system sizes by employing an MPS ansatz where we represent the Hilbert space with a six dimensional local basis. (See App.~\ref{app:MPS} for details.)}. For a single spin configuration $\mathbf{s}_\textsc{x}$, the state of the quantum walk at time $t$ can accordingly be written in terms of the eigenstates of $\hat{W}_{\mathbf{s}_\textsc{x}}$ such as  
\begin{equation}
\ket{\Psi_{t}^{\mathbf{s}_\textsc{x}}}=\sum_k^{2N} b_k^{\mathbf{s}_\textsc{x}}(t)\ket{\xi_{k}^{\mathbf{s}_\textsc{x}}}\otimes\ket{\mathbf{s}_\textsc{x}},
\label{eq:totSt}
\end{equation}
where $\ket{\xi_{k}^{\mathbf{s}_\textsc{x}}}$ is the $k^\mathrm{th}$ eigenstate with amplitude $b_k^{\mathbf{s}_\textsc{x}}(t)$. For example, the two completely $\hat{X}$-polarized spin configurations, where all $s_n$ are either $+$ or $-$, yield the standard quantum walk with all of its eigenstates $\ket{\xi_{k}^{\mathbf{s}_\textsc{x}}}$ extended. On the other hand, it is known that the existence of a single spin-flip disorder leads some localized eigenstates  around this impurity~\cite{danaci2019non,wojcik2012trap}. In general, when the distribution of $s_n$ are disordered, the eigenstates can become localized. 

For the dynamics to be localized, the delocalized eigenstates should not effectively contribute to the dynamics, which naturally depends on the choice of the initial state. We will concentrate on an initial state with all spins polarized in $\hat{Z}$, such as $\mathbf{s}_\textsc{z}=\left(00\hdots0\right)$, which is an equal weight superposition of all $\ket{\mathbf{s}_\textsc{x}}$ eigenstates. Therefore, the time evolution occurs independently in all spin sectors $\{\mathbf{s}_\textsc{x}\}$ so that
\begin{equation}
\ket{\Psi_t} = \hat{W}_M^{t}\ket{\Psi_0}=\frac{1}{2^{N/2}}\sum_{\mathbf{s}_\textsc{x}}\left\{\hat{W}_{\mathbf{s}_\textsc{x}}^t\ket{\psi_0}\otimes\ket{\mathbf{s}_\textsc{x}}\right\}.
\label{eq:evInEsBasis}
\end{equation}
Noting that the projection operator $\hat{P}_n=I_2\otimes\ket{n}\bra{n}\otimes I_{2^N}$ does not mix the different spin states, the probability distribution will be equivalently given by an average of the disordered walks $\hat{W}_{\mathbf{s}_\textsc{x}}$ as
\begin{equation}
P_n(t)=\frac{1}{2^{N}}\sum_{c,\mathbf{s}_\textsc{x}} 
    \left\vert \bra{c,n} W_{\mathbf{s}_\textsc{x}}^t\ket{\psi_0} \right\vert^2.
\label{eq:prDist}
\end{equation}
The probability $P_n(t)$ becomes an average of walks over the different binary distribution of local phases $e^{\pm i\phi}$.
Hence, if walks under $\hat{W}_{\textbf{s}_\textsc{x}}$ are localized on average, the dynamics will be too. We obtain the probability distribution of the quantum walker by employing a numerically exact MPS ansatz which evaluates Eq.~\eqref{eq:totSt} by truncating the state for a given precision in all spin configurations (see App.~\ref{app:MPS} for details). 
\bnd{}{Performing the averaging in Eq.\eqref{eq:prDist} for the walker through the total quantum evolution is technically equivalent to the proposition by Paredes et al.~\cite{paredes2005exploiting}.
In that context simulating a quantum system with classical random variables is accomplished by exploiting quantum parallelism using an auxiliary quantum system. 
The auxiliary system which corresponds to the local spins in our model does not have self-dynamics.}

\bnd{Moreover, by considering this MPS results to be our reference, we show that the computational complexity of numerically calculating $P_n(t)$ for the unitary $\hat{W}_M$ can be reduced to that of $\hat{W}_{\mathbf{s}_\textsc{x}}$ by random sampling Eq.~\eqref{eq:prDist} as we present in Sec.~\ref{sec:locEntDec}.}{Having shown that the computational complexity of numerically calculating $P_n(t)$ for the unitary $\hat{W}_M$ can be reduced to that of $\hat{W}_{\mathbf{s}_\textsc{x}}$, we also obtain the walker's probability distribution by random sampling via Eq.~\eqref{eq:prDist}}. 
 Statistically, the finite number of random spin configuration samples are most likely to be chosen among the vicinity of zero total $X$-polarization. The agreement between our sampling and MPS results confirms that those configurations are the ones that contribute to the dynamics significantly as we present in Sec.~\ref{sec:locEntDec}. 

\begin{figure}[t]
\centering
\includegraphics[scale=0.85]{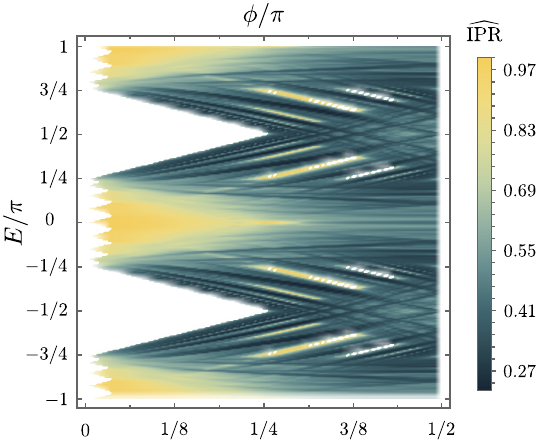}
\caption{\label{fig:enSpec}Quasi-energy $E$ spectrum and the inverse participation ratio ($\widehat{\text{IPR}}$) of the corresponding eigenstates with respect to the interaction parameter $\phi$ for the quantum walker stepping with $\hat{W}_M$. For a given $E$, the inverse participation values are averaged over degenerate cases and normalized by the number of spins $N=18$. After the critical spin rotation angle $\phi~=~\pi/4$, the band gaps close and the quasi-energy eigenstates become localized.}
\end{figure}

Having reduced the problem to that of a disordered walker, we emphasize that quantum walks are known to exhibit Anderson localization in the presence of position-dependent static disorder (time-independent), which has been demonstrated analytically~\cite{joye2010,ahlbrecht2011disordered,hamza2009dynamical}, numerically~\cite{nicola2014,pankov2019anderson}, and experimentally~\cite{crespi2013anderson,schreiber2011}. 
We show that when $\phi$ is sufficiently large, similar to one-dimensional Hamiltonian models with disorder the probability distribution 
around the origin stays exponentially localized showing the striking signature of Anderson localization. 
For finite system sizes, the transition between spreading and localized regimes has also been shown~\cite{schreiber2011}.
Finally, we note that the parameter $\phi$ in the unitary evolution related to disorder in our model yields dynamical localization, but should not be directly associated with the random potential of Hamiltonian models. 
 
\subsection{Figures of Merit\label{sec:figOfMer}} 

To quantify the localization of the quantum walker, we will employ two measures derived from the probability distribution of the quantum walker, namely, the variance 
\begin{equation}
\sigma^2_t=\sum_n n^2 P_n(t)
\end{equation}
and the normalized Inverse Participation Ratio
\begin{equation}
  \widehat{\text{IPR}}=\left(N \sum_{n=1}^{N} P_n^2\right)^{-1}. 
\label{eq:ipr}
\end{equation}
The variance of the probability distribution is a well-known measure for classifying the spreading rate of the quantum walker. While a ballistic spread as in the standard disorder-free quantum walk case gives a quadratic growth in the variance $\left(\sigma^2_t \sim t^2\right)$, a classical random walk is diffusive with a variance that increases linearly $\left(\sigma^2_t \sim t\right)$ with respect to the step number. Apart from these, a localized probability distribution yields either a constant or a fluctuating variance which has no overall increase in time. The inverse participation ratio, on the other hand, estimates the average number of sites where the quantum walker is spread over uniformly. We employ the inverse participation ratio which is normalized over the size of the lattice $N$ as given in Eq.~\eqref{eq:ipr}. Therefore, $\widehat{\text{IPR}}\rightarrow 1$ if the walker is spread over the position space uniformly and $\widehat{\text{IPR}}\rightarrow 0$ if the walker is localized on a single site, as $N\rightarrow\infty$. 

After we discuss the localization properties of the quantum walker, we will turn our attention to the spin chain itself and examine the entanglement entropy between its bipartite subdivisions for investigating further properties of DFL. Let $A$ and $B$ represent subsystems forming the two halves of the total system.~\footnote{\bnd{As the walker degrees of freedom involve all sites, we can either combine the walker subsystem with the left or right half of spins or also spatially divide its Hilbert space into two by introducing a vacuum state with no quantum walker.}{In addition to the local spins found in each partition, we split the position and coin degrees of freedom of the walker to encompass the sites corresponding to partitions A and B, separately. Therefore, we also consider a vacuum state representing the absence of the walker at the given partition (please also see the local bases for the MPS representation in the Appendix A). The Hilbert space of a partition 
consists of the tensor product of all the spin states in that partition and all the states of the walker in that partition including its vacuum state. For $N=N_A+N_B$, the number of degrees of freedom in partition $A$ is given by $(2N_A+1)2^{N_A}$. 
If the walker state is also traced out, then the spin partition has $2^{N_{A}}$ states. }} The reduced density matrix of system $A$ is obtained by tracing over system $B$ as $\rho_A(t)=\tr_B\outerp{\Psi_t}{\Psi_t}$ where $\ket{\Psi_t}$ is the total state of ($A+B$) given in Eq.~\eqref{eq:totSt}. Since $\ket{\Psi_t}$ is a pure state by assumption, we employ the von Neumann entropy 
\begin{equation}
S(\rho_A)=-\tr\left[\rho_A\log\rho_A\right]
\label{eq:EE}
\end{equation}
to measure the entanglement between system $A$ and $B$. The calculation of Eq.~\eqref{eq:EE} as a function of $t$ and walker-spin interaction parameter $\phi$ is a difficult problem to handle due to the large Hilbert space in question. Nevertheless, MPS ansatz method provides an efficient way of calculating Eq.~\eqref{eq:EE} as we will later elaborate in App.~\ref{app:MPS}.

\section{\label{sec:results}Results}

\subsection{Energy spectrum}

The dynamics of quantum walks can be regarded as a stroboscopic realization of an effective Hamiltonian defined as $\hat{W}_M \equiv e^{-i\hat{H}_\text{eff}}$. Similar to quasi-momentum in systems with discrete translational symmetry, the eigenvalues of $\hat{H}_\text{eff}$ are called quasi-energies since $\hat{W}_M$ is applied periodically, i.e, the system possesses a discrete time-translational symmetry~\cite{kitagawa2012}. We start by discussing the quasi-energy spectrum of the step operator $\hat{W}_M$ (the eigenvalues of $\hat{H}_\text{eff}$) as a function of $\phi$ as shown in Fig.~\ref{fig:enSpec} for a system size of $N=18$. We note here that our calculations do not yield any change in quasi-energy spectra after $N\geq15$. The surface plot has quasi-energy $E$ on the vertical axis and each eigenstate is colored with its IPR value. At $\phi=0$ the system has a $2^{18}$-fold degenerate disorder free quantum walk spectrum consisting of two bands.  The bands are separated by two band gaps of width \bnd{$\pi/2$}{$2\theta = \pi/2$} and all states are delocalized for the quantum walker indicated by the light color map. As the coupling is turned on, localized states indicated by darker colors appear first at the edges of the two bands whereas delocalized states are mostly in the middle of the bands. At $\phi=\pi/4$ band gaps are closed and almost all eigenstates are localized for $\phi>\pi/4$. One cannot directly identify a mobility edge due to the fact that quasi-energies from different spin sectors get also shifted depending on the $\phi$ value. For example, the $s_n=-\isk{1}{}$ and $+\isk{1}{}$ (for all $n$) spin sectors always have delocalized states, however their quasi-energies are shifted by $\pm \phi$ with respect to the standard walk spectrum \bnd{}{causing the band gap to close at $\phi=\theta =\pi/4$}. These shifts explain delocalized states (indicated by light colors) in the spectrum seen for $\phi > \pi/4$ and are also the cause of the upward and downward moving branches for localized states. Fig.~\ref{fig:enSpec} therefore visualizes the single-particle localization as it appears in combination from the different spin sectors $\mathbf{s}_\textsc{x}$. As discussed in Sec.~\ref{sec:DFL}, the localized eigenstates are responsible for the emergent localization which is presented in the following section.

\subsection{\label{sec:locEntDec} Localization, entanglement and decoherence}

\begin{figure}[t]
\centering
\includegraphics[scale=0.60]{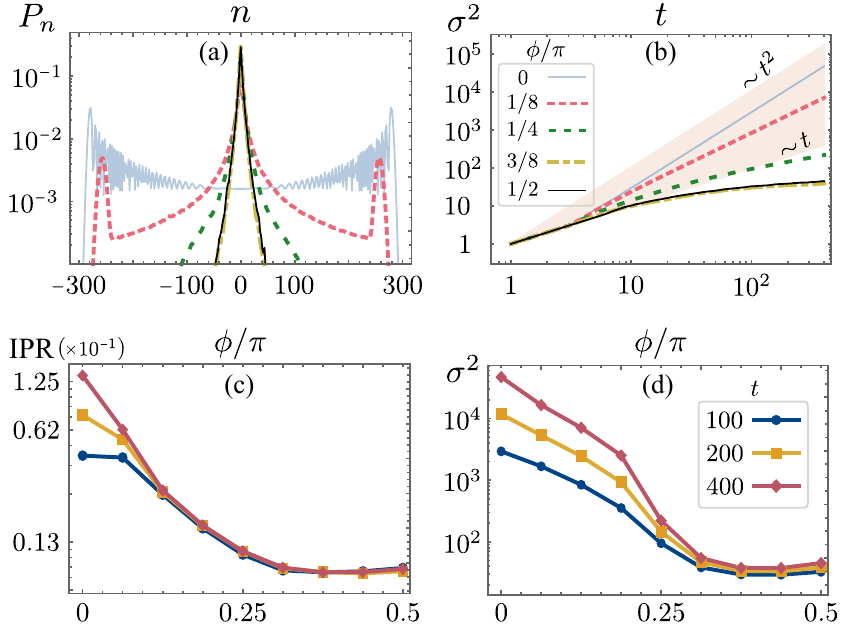}
\caption{\label{fig:fig3}Probability distribution of the walk in the position space after 400 steps (a) and its variance in position space as a function of step number (b), for the initial state $\ket{\Psi_0}$. Inverse participation ratio (IPR) (c) and variance (d) are shown as a function of the spin rotation angle $\phi$, at different time steps. IPR is normalized by the lattice size $N~=~801$. The walker is exponentially localized for $\phi>\pi/4$.} 
\end{figure}

In this section, we consider the dynamical properties of the quantum walker which is initially localized at $n=0$. The initial spin state is chosen to be a product of $\hat{Z}$-polarized local spins $\ket{\mathbf{s}_\textsc{z}}$ such that
\begin{equation}
\ket{\Psi_0} = \ket{\chi_0}\otimes\ket{n=0}\otimes \ket{0}^{\otimes N}
\label{eq:iniSt}
\end{equation}
with the coin state $\ket{\chi_0}$ yielding a symmetric distribution around the origin as mentioned in Sec.~\ref{sec:walk}. The initial spin state $\ket{\mathbf{s}_\textsc{z}} \equiv \ket{0}^{\otimes N}$ is translationally invariant.
This choice does not restrict the applicability of our results for any $\ket{\mathbf{s}_\textsc{z}}$ as discussed in Sec.~\ref{sec:DFL}.

Here we present the results calculated with the random sampling method mentioned in Sec.~\ref{sec:DFL}. We find that an average over a few thousand samples are in perfect agreement with MPS calculation results, which we performed up to $t=100$. With $4000$ $\mathbf{s}_\textsc{x}$ samples, the quantum walker's probability distribution in position space after $t=400$ time steps is shown in Fig.~\ref{fig:fig3}(a) for increasing values of the interaction parameter $\phi$. The disorder free standard quantum walk ($\phi=0$) displays ballistic peaks and the variance of the probability distribution increases as $\sigma^2\sim t^2$ which gives a slope of two on the log-log plot shown in Fig.~\ref{fig:fig3}(b). As the coupling to the spins is turned on ($\phi\neq 0$), the quantum walker remains partially localized near the $n=0$ with less pronounced side peaks. Note that the interference effects leading to oscillations in the probability of the standard walk are wiped out for $\phi>0$ as seen in Fig.~\ref{fig:fig3}(a). This is due to decoherence effects induced by the spin environment. For $\phi=\pi/8$, the spread of the ballistic tails decreases as can be seen from the reduced slope of $\sigma^2$. Furthermore, at the critical value $\phi=\pi/4$, the tails become completely suppressed and $\sigma^2$ displays a near diffusive ($\sigma^2\sim t$) behavior. The shaded triangle in Fig.~\ref{fig:fig3}(b) highlights the range between the ballistic spreading and the diffusive limit with its upper and lower edges, respectively. The walker remains exponentially localized for $\phi>\pi/4$ with a localization length of $\lambda \sim 1.6$ which we calculate by fitting $\exp(-2|n|/\lambda)$ to the probability distribution $P_n$. Note that the localization length is on the order of the lattice spacing. In this regime, the spread of the quantum walker is sub-diffusive and the variance approaches a constant value as $t$ increases. 

The change in the probability distribution as a function of $\phi$ shows the crossover from ballistic spreading to a complete localization in our disorder free model. Fig.~\ref{fig:fig3}(c) and Fig.~\ref{fig:fig3}(d) show the inverse participation ratio and the variance as a function of $\phi$ for different step numbers, respectively. These plots indicate that regardless of how many steps were taken, spreading of quantum walker is small and it remains constant for $\phi > \pi/4$. The gradual decrease in IPR and $\sigma^2$ as $\phi$ goes from $0$ to $\pi/4$ results from the suppression of the side peaks and their decreasing spreading rate, respectively.

\begin{figure}[t]
\centering
\includegraphics[scale=0.54]{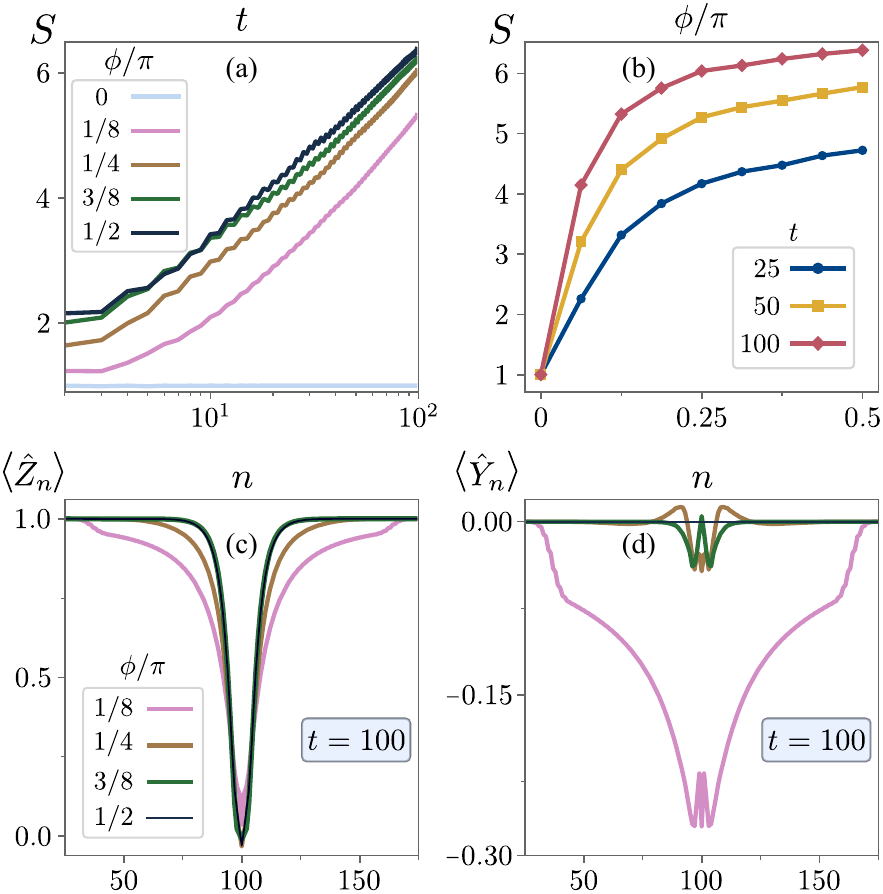}
\caption{\label{fig:fig4}Entanglement entropy between two halves of the spin chain as a function of time (a) and as a function of the angle $\phi$ at different time steps (b). Spin expectations along z- (c) and y-axes (d).}
\end{figure}

The step operator $\hat{W}_M$ creates entanglement between the position, coin, and spin degrees of freedom which can be quantified by the von Neumann entropy of subsystems as we mentioned in Sec.~\ref{sec:figOfMer}. We calculate this entanglement entropy $S$ of one partition by spatially dividing the degrees of freedom in half. This can easily be accessed via the singular values in the MPS simulations (see App.~\ref{app:MPS}) and is plotted as a function of time in Fig.~\ref{fig:fig4}(a) for different values of the coupling $\phi$. Even though the growth is slower for partially localized cases ($\phi<\pi/4$), the entanglement shows sublinear growth starting with the second decade of the time evolution for all non-zero values of $\phi$. (For the standard quantum walk at $\phi=0$ the entanglement saturates at $\log_2 2=1$ (horizontal blue line) with the initial state spreading equally into the left and rights half of the system.) The sublinear growth happens faster with increasing $\phi$ as seen in Fig.~\ref{fig:fig4}(b) for three different times and $S$ does not change significantly after $\phi\geq\pi/4$ once the system is completely localized. We emphasize that the behavior of $S$ is \bnd{characteristic of}{can be observed in} other localized systems such as MBL and DFL. Here, the sublinear growth is also related to the walker's Hilbert space being small and the generation of entanglement in the spin chain has to occur via interactions with the small Hilbert space of the walker. The long-time behavior of $S$ is therefore similar for all $\phi$ even when the walker is partially localized for $\phi<\pi/4$. \bnd{}{We present a more detailed analysis of the entanglement spread on the lattice as a function of time in App.~\ref{app:MPS}.}

Finally, we consider the spin sub-system results for which MPS calculations are again required. The $\hat{X}_n$ expectation values of local spins remain zero throughout the evolution as they constitute the conserved quantities in the model. The $y$-polarization also vanishes for the initial state $\ket{\psi_0}$ since the spins are polarized in $z$-direction. During the evolution, the spin expectation values decohere because of the interaction with the walker. Therefore, the spatial localization in $\langle \hat{Z}_n \rangle$ appears with a similar structure to that of the walker and is shown in Fig.~\ref{fig:fig4}(c) at $t=100$. The expectation values $\langle \hat{Y}_n \rangle$ in Fig.~\ref{fig:fig4}(d) similarly show the spreading of the side peaks for $\phi=\pi/8$ and the spin textures remain localized for $\phi\geq \pi/4$. We note that differently from the localization of the walker's probability distribution, the spin textures show a dependence on the initial coin state $\ket{\chi_0}$. An imbalance in the weights of $\ket{s_n=\pm \isk{1}{}}$ states changes the symmetric $\langle \hat{Z}_n \rangle$ distribution and the phase difference effects the $\langle \hat{Y}_n \rangle$ distribution. 

\section{\label{sec:conclusion}Discussion and conclusion}

We have studied disorder-free localization of a quantum walker coupled to local spins on a one-dimensional lattice. Similar to models of quantum walks on graphs coupled to spins living on nodes~\cite{camilleri2014quantum} or links~\cite{verga2020}, the local spins in our model live on lattice sites and do not act on the coin. The Hamiltonian and the initial spin state we have chosen are translationally invariant, yet localization occurs merely due to the interactions between the walker and the local spins. Therefore, our model exhibits a discrete-time version of the DFL introduced recently~\cite{smith2017disorder} in a periodically driven system. One of the advantages of our model is the presence of the extensive number of conserved moments which provides a computationally easier method to analyze the spread of the walk. Due to the conserved quantities, the evolution problem reduces to that of disordered walks. Since we chose the initial state to be a superposition of all possible spin disorder configurations, the spatial probability of the walker can be obtained through an ensemble averaging over all possible spin configurations. 

We have observed two regimes of localization depending on the strength of the interaction between the walker and the local spins.  
\bnd{When the interaction is weak, only some of the eigenstates are localized around the origin. }{Unlike the conventional Anderson insulators in one-dimension, , where all the eigenstates are localized for arbitrarily small disorder strengths, for weak interactions delocalized eigenstates exist in the quasi-energy spectrum.}
For weak coupling, similar to the mobility edges found in three-dimensional Anderson localization~\cite{Semeghini2015} and many-body localized systems under weak disorder~\cite{abanin2017}, the localized states are concentrated near the band edges of the spectrum. As a result, the walker gets partially localized around the origin~\cite{prokof2006decoherence} and the ballistic tails are suppressed. 
The oscillations in the probability distribution become smooth due to the decoherence effects with the spin environment. 
\bnd{The energy spectrum for weak interaction is different from those of the one-dimensional Anderson insulators, where all the eigenstates are localized for arbitrarily small disorder strengths. }{}
As the interaction strength increases, the band gap gets smaller and the eigenstates become more localized. 
\bnd{Above a critical strength}{As the coupling increases}, the band gap closes and the ballistic tails disappear with the walker being \bnd{exponentially}{completely} localized. The localization length is reduced down to the order of the lattice spacing and does not vary appreciably as a function of the interaction strength above the critical value. \bnd{}{General wisdom is that one-dimensional systems exhibit Anderson localization under continuous-time evolution for arbitrarily small disorder. Exceptional cases usually possess correlated disorder or long-range interactions. 
The emergent disorder in our model is uncorrelated disorder. The motion of the quantum walker and the interaction between the walker and the local spins are short ranged.
Therefore, with our intermediate time and corresponding system size simulations we cannot rule out the possibility of finite size effects. 
On the other hand, 
in contrast to the Hamiltonian models, the discrete-time dynamics here is given by a unitary step operator. 
There are also suggested models of many-body localization where stroboscopic dynamics such as driving by an external light may cause delocalization \cite{bordia2017periodically,lenarvcivc2018activating,Ray_2018,Ray2_2018}.
\newline
We think that the apparent transition around $\phi = \pi / 4$ in our model can also be related to the natural length scale in the model which is the lattice spacing. A possible explanation of the results for $\phi < \pi / 4$ could be that the Anderson localization length is larger or comparable to our system size, thus the finite size effect mentioned above would be relevant. For $\phi > \pi / 4$, the conclusion is that the Anderson localization length becomes comparable to the lattice spacing. Therefore, the change of the localization length with $\phi$ could be the reason behind the different behaviour. This is in agreement with the sharp features in the quasi-energy spectrum in the sense that for $\phi > \pi / 4$ there is no band gap and we find localized states for almost all quasi-energy values.}

We have also introduced a straightforward implementation of the MPS ansatz on the lattice which is essential to study the spin dynamics and the entanglement entropy of different partitions of the spin chain. The discrete-time unitary evolution can be implemented without any formal approximations. \bnd{Our results show that for all the values of the interaction strength, the growth of the entanglement entropy is logarithmically slow. The growth rate increases as the interaction gets stronger and it saturates when the walker is exponentially localized. The logarithmic suppression of the entanglement growth is similar to the DFL and MBL systems previously studied in the literature.}{Our results show that the growth of the entanglement entropy depends on the interaction strength. In the intermediate time scale of our simulations, we observed that for small interaction strengths, the partial localization of the walker yields linear spread of the entanglement towards edges combined with sublinear spread around the origin. As the interaction gets stronger, linear spread is suppressed while the sublinear spread prevails and causes the growth rate of the entanglement to increase. For strong coupling no ballitic spread is observed and the growth rate saturates when the localization length becomes comparable to the lattice spacing. The sublinear suppression of the entanglement growth is similar to some of the DFL and MBL systems previously studied in the literature.}

\isk{We have tested}{In App.~\ref{app:field}, we further test the} robustness of the quantum walker's localization and spin expectation values by applying a symmetry-breaking uniform field to local spins. With a small field along $z$-direction, we show that the local $x$-polarization is not conserved any more and the results change over to those of a classical random walk. We estimate the time scale for this change. 

Our work connects quantum walk models as examples of periodically driven systems to the recent studies on disorder-free localization and related areas. The use of MPS and tensor network states in general can lead to further investigations of quantum walks interacting with local degrees of freedom. A natural extension of our study would be to consider a quantum walk on a higher dimensionsional local spins where a transition to classical random walk and a diffusive spread is expected. Furthermore, the fate of localization in two and three dimensional spin lattices can be studied. It would also be interesting to analyze the connection between localization and non-Markovianity in similar models.

Lastly, it is worthy to note that our sampling results are experimentally feasible with current experimental technology. The electro-optical modulators (EOMs) presented in the time-multiplexing quantum walk scheme make it possible to introduce position-dependent random phases for the walker~\cite{schreiber2010,schreiber2011,elster2015,nitsche2016,nitsche2018}. EOMs affect the light pulses (walker) going through a fiber loop in the time domain representing the position space. Therefore, appropriate programming for EOMs may allow the realization of any binary disordered landscape presented in our model, and the localization results could be obtained by averaging over different realizations.

\begin{acknowledgments}
B. D. thanks Juan Jos\'{e} Garc\'{i}a Ripoll for useful discussions on matrix product states. \.{I}. Y. thanks A. G\'abris for helpful discussions and the support by M\v{S}MT under Grant No. RVO 14000 and the Czech Science Foundation under Grant No. GA CR 19-15744Y. A. L. S. benefited from the 2015 Ghent summer school on tensor networks and acknowledges the hospitality of CNLS. G. K. is supported by the BAGEP Award of the Science Academy, the TUBA-GEBIP Award of the Turkish Academy of Sciences, and also by the Technological Research Council of Turkey (TUBITAK) under Grant No. 117F317. B. \c{C}. is supported by the BAGEP Award of the Science Academy, the Technological Research Council of Turkey (TUBITAK) under Grant No. 117F317, and also by the Research Fund of Bah\c{c}e\c{s}ehir University (BAUBAP) under project no: BAP.2019.02.03. 
\end{acknowledgments}

\appendix
\counterwithin{figure}{section}
\renewcommand\thefigure{\thesection.\arabic{figure}}    
\setcounter{figure}{0}

\section{\label{app:MPS}Matrix Product States Formalism}

In matrix product states (MPS) formalism, different degrees of freedom and local interactions between them are represented by local tensors. The general form of a matrix product state can be written as~\cite{schollwock2011density}
\begin{equation}
\ket{\Psi} = \sum_{\mathbf{q}} \text{Tr} \left[ A^{q_1} \dots A^{q_N} \right]
\ket{\mathbf{q}} 
\label{eq:mps}
\end{equation}
where $q_n$ are the local degrees of freedom and $\ket{\mathbf{q}} \equiv \ket{q_1, \dots, q_N}$ form a basis for the global Hilbert space. Each local tensor $A$ has three indices corresponding to one physical dimension and two bond dimensions. The physical dimension is determined by the number of local physical degrees of freedom ($q_n$) associated with each tensor. The coefficients of the wave function are calculated as a matrix multiplication by contracting the virtual bond indices. The bond dimension between the tensors $A^{q_n}$ and $A^{q_{n+1}}$ is obtained by applying singular value decomposition for the partitioning of the matrix product at the $n^\mathrm{th}$ bond. The number of non-zero singular values $\{\sigma_\alpha\}$, is equal to the bond dimension $D$. Using the singular value decomposition, the tensors can be brought to canonical form. In this form, matrices are rewritten in right or left orthonormal bases. This actually provides an efficient way of manipulating the MPS state and computing the desired expectation values. For example, on-site expectation values can be calculated using the canonical form pivoted on that site as the tensors to the left and right of the site reduce to identities when they are contracted.

The squares of the singular values are the eigenvalues of the reduced density matrix of either subsystem, i.e. the partition to the left or right of a given bond. Therefore, the von Neuman entropy (see Eq.~\eqref{eq:EE}) becomes 
\begin{equation}
S =-\sum_{\alpha=1}^D \left[\sigma_{\alpha}^2 \log \sigma_{\alpha}^2 \right],
\label{eq:EEmps}
\end{equation}
and the eigenvalues also satisfy $\sum_{\alpha} \sigma_\alpha^2 = 1$.
For a product state, the bond dimensions of all the local tensors are equal to one (\isk{$\sigma_\alpha=1$}{$\sigma_1=1\, \forall\, n$)} which means zero entanglement entropy. As the entanglement between two parts of the system increases, the dimension of the bond connecting the two parts increases as well. As the Hilbert space dimension of a many-body system scales exponentially with system size, a random quantum state will likely have large bond dimensions.
The efficiency of the MPS representation relies on the fact that low-energy states of one-dimensional physical systems with short-range interactions generally have low entanglement~\cite{orus2014practical}, which in turn keeps the bond dimensions of the local tensors bounded. In practice, bond dimension can be restricted by imposing an upper limit, or eliminating singular values below a truncation tolerance, $\delta$. 

In conventional MPS algorithms, a given Hamiltonian is exponentiated to obtain the evolution operator. Due to the commutation errors, approximate methods such as Suzuki-Trotter expansion~\cite{suzuki1990fractal,suzuki1991general} are used in numerical calculations, in general. 
Since for the discrete-time quantum walk the unitary evolution operator is defined instead of the Hamiltonian, we do not need a Suzuki-Trotter expansion for numerical MPS simulations. The implementation below is numerically exact for the precision determined by the truncation tolerance.

For a single spinless particle moving on a lattice, a straightforward MPS representation is to associate a basis composed of two states for each lattice site. These states correspond to the vacuum and particle (being present on the site) states. (Note that this is not the most efficient representation because the MPS ansatz spans a $2^N$ dimensional physical Hilbert space including the no particle vacuum state as well as many-particle states with a maximum of one particle per site.~\cite{Delgado1999})
For a quantum walk, the local basis can be extended to include the coin degrees of freedom such that
\begin{align}
	\ket{q_n=0} &\leftrightarrow \ket{\mathrm{vacuum},n}, \nonumber \\
	\ket{q_n=1} &\leftrightarrow  \ket{c=0,n}, \\
	\ket{q_n=2} &\leftrightarrow  \ket{c=1,n} \nonumber
\end{align}
forming a three-dimensional basis and $\ket{c,n}$ are quantum walker states defined in Sec.~\ref{sec:walk}. 

\begin{figure}[t]
\centering
\includegraphics[scale=0.4]{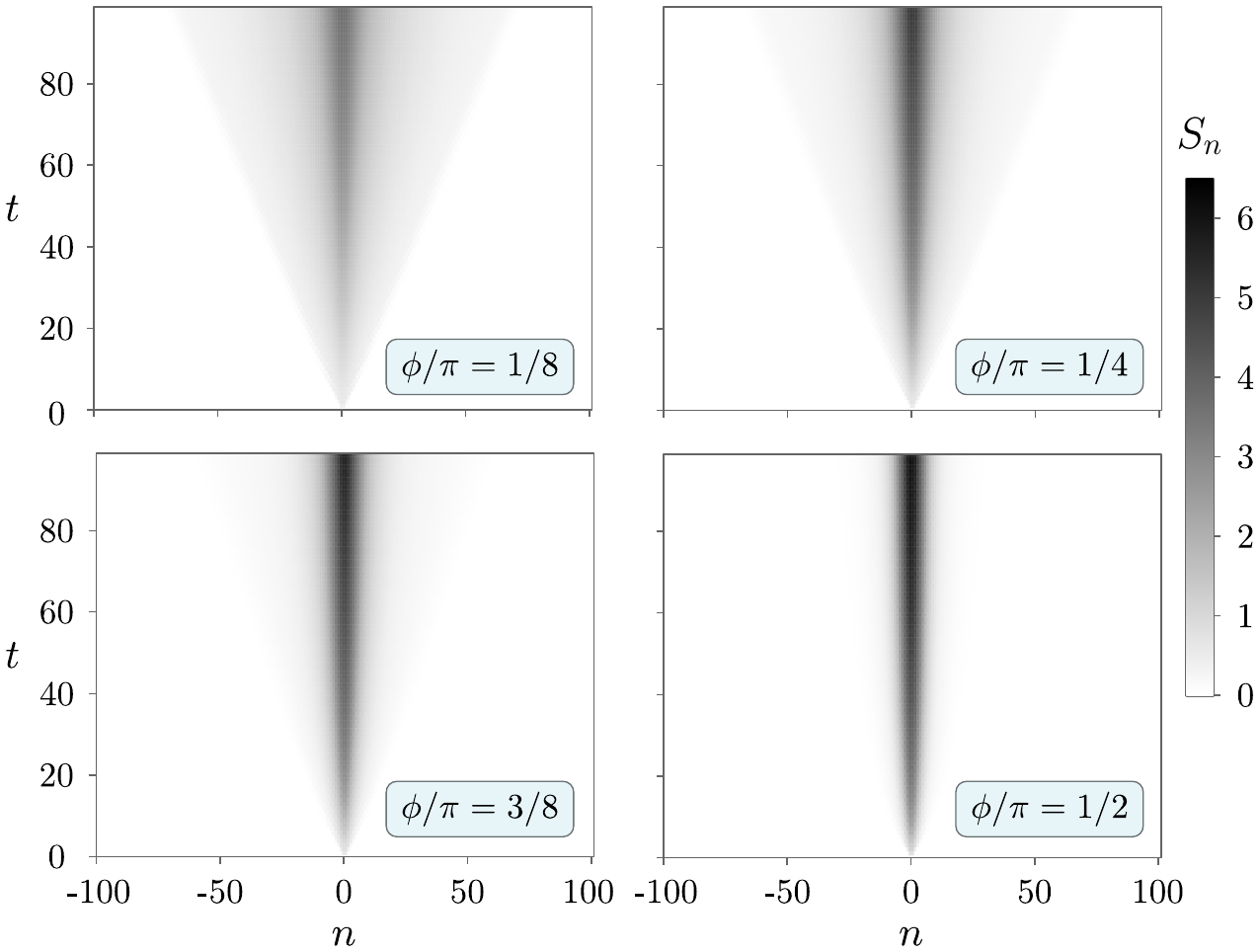}
\caption{\label{fig:A1new} Entanglement entropy between bi-partite partitions of the chain at the $n^{\text{th}}$ site, $S_n$, as a function of time.}
\end{figure}

The unitary walk operator $\hat{W}$ consists of the successive application of the coin $\hat{C}$ and shift $\hat{T}$ operators. With the above identification of the local basis, $\hat{C}$ can be implemented as
\begin{equation}
	\mathbf{I}_1 \oplus \exp{\left(-i\theta\mathbf{X}\right)} \leftrightarrow \hat{C}
\end{equation}
where the matrix representation on the left is contracted with the physical index $q_n$ of the local tensors at each site $n$.

The conditional shift operator $\hat{T}$ can be broken into the left $\hat{T}_\mathrm{L}=\sum_n \outerp{c=1,n}{c=1,n+1}$ and the right $\hat{T}_\mathrm{R}=\sum_n \outerp{c=0,n+1}{c=0,n}$ shift operators, which can be implemented by the two-site application of the matrices
\small
\begin{equation} 
	\begin{pmatrix}
		1 & 0 & 0 \\
		0 & 0 & 0 \\
		0 & 0 & 1
	\end{pmatrix}_n
	\!\! \otimes
	\begin{pmatrix}
		1 & 0 & 0 \\
		0 & 1 & 0 \\
		0 & 0 & 1
	\end{pmatrix}_{n+1} \!\!\! +
	\begin{pmatrix}
		0 & 1 & 0 \\
		0 & 0 & 0 \\
		0 & 0 & 0
	\end{pmatrix}_n
	\!\! \otimes
	\begin{pmatrix}
		0 & 0 & 0 \\
		1 & 0 & 0 \\
		0 & 0 & 0
	\end{pmatrix}_{n+1}
	\!\!\! \leftrightarrow \hat{T}_\mathrm{R}
\end{equation}
\begin{equation}
	\begin{pmatrix}
		1 & 0 & 0 \\
		0 & 1 & 0 \\
		0 & 0 & 1
	\end{pmatrix}_{n-1}
	\!\!\!\!\! \otimes
	\begin{pmatrix}
		1 & 0 & 0 \\
		0 & 1 & 0 \\
		0 & 0 & 0
	\end{pmatrix}_{n}
	\!\!\! +
	\begin{pmatrix}
		0 & 0 & 0 \\
		0 & 0 & 0 \\
		1 & 0 & 0
	\end{pmatrix}_{n-1}
	\!\!\!\!\! \otimes
	\begin{pmatrix}
		0 & 0 & 1 \\
		0 & 0 & 0 \\
		0 & 0 & 0
	\end{pmatrix}_{n}
	\!\!\! \leftrightarrow \hat{T}_\mathrm{L}
\end{equation}
\normalsize
which are contracted with physical indices of sites $n$ and $n+1$. As $\hat{T}_\mathrm{L}$ and $\hat{T}_\mathrm{R}$ act on different coin states, they commute with each other. Thus, the order of their application does not matter. However, for both of them, the even- and odd-bond hopping terms do not commute with each other. To avoid commutation errors, when sweeping over consecutive lattice bonds from left to right, $\hat{T}_\mathrm{L}$ can be applied. Similarly, when the sweep direction is switched $\hat{T}_\mathrm{R}$ can be applied \isk{}{without introducing a numerical error.}~\footnote{Note that in Suzuki-Trotter expansion, the error due to the non-commuting terms in the Hamiltonian scales as a power of the time-step, where the power is determined by the order of expansion. Therefore, time-step should be small to reduce the error, which increases simulation time in general. Here, the size of the time-step does not have an effect on the error.}.

The local spins can be included by extending the local basis as a direct product with that of the local spin $\ket{s_n}$ giving a $3\times 2=6$ dimensional local Hilbert space. (For a single walker, the bond arrangement is such that non-zero amplitudes only appear for states with $q_n = 0$ for all $n$ except one.)
Finally, the matrix-product-operator for the interaction $\hat{M}$ can be implemented with the following matrix
\begin{equation}
	\left[ \mathbf{1}_1 \oplus \mathbf{0}_2 \right] \otimes\mathbf{I}_2 +
	\left[ \mathbf{0}_1 \oplus \mathbf{I}_2 \right] \otimes\exp{\left(-i\phi\mathbf{X}\right)} \leftrightarrow \hat{M},
\end{equation}
which is again to be applied at every site $n$ to the tensor $A^{q_ns_n}_{b_{n-1},b_n}$ with bond indices $b_{n-1}$ and $b_n$ to the left and right, respectively. In fact, the combination $\hat{C}\hat{M}$ can be performed together. Each step of $\hat{W}_M = \hat{T}\hat{C}\hat{M}$ is composed of consecutive application of the on-site coin and interaction operators and a single sweep of the right and left translation operators. 

To compute the spin expectations, we extend the Pauli operators to include the coin space. For instance, Pauli X operator, $\hat{X}$, is replaced with
\begin{equation}
	\left[ \mathbf{I}_1 \oplus \mathbf{I}_2 \right] \otimes \mathbf{X} \leftrightarrow \hat{X}_n,
\end{equation}
If the state is in canonical form with respect to site $n$, then computing the expectation $\langle \hat{X}_n \rangle$ reduces to the contraction of the above operator with the local tensor and its Hermitian conjugate, i.e., 
\begin{equation}
	\langle \hat{X}_n \rangle =\!\!\!\!\!\! \sum_{\substack{q_ns_n, q'_ns'_n, \\b_n,b_{n-1}}} \!\!\!\!\!\! \left( A^{q'_ns'_n}_{b_{n-1},b_n} \right)^* \left[ \left[ \mathbf{I}_1 \oplus \mathbf{I}_2 \right] \otimes \mathbf{X}\right]_{q'_ns'_n,q_ns_n}  A^{q_ns_n}_{b_{n-1},b_n} \, .
\end{equation}

\begin{figure}[t]
\centering
\includegraphics[scale=0.53]{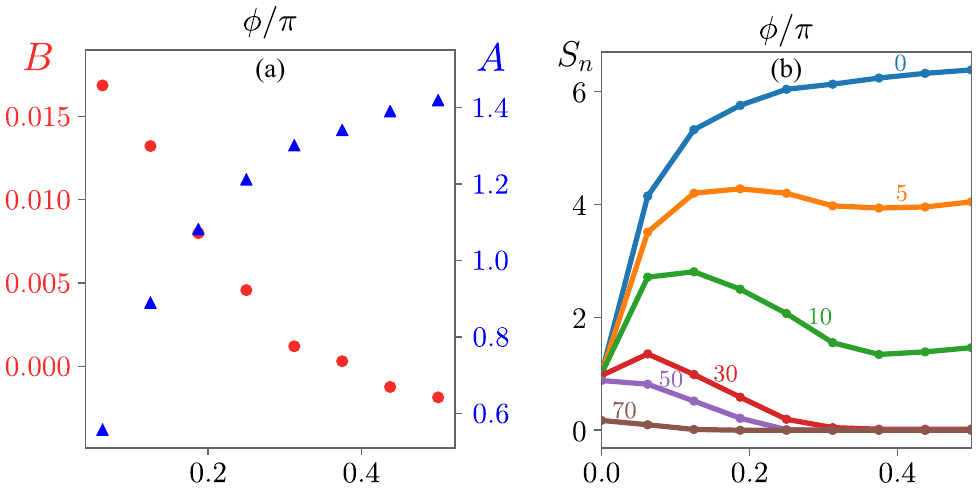}
\caption{\label{fig:eEnt} (a) The logarithmic, $A$, and the linear, $B$, contributions to the growth of the entanglement entropy, $S$ (see Eq.\eqref{eq:entanglement_vstime}). (b) Entanglement entropy between bi-partite partitions of the chain at the $n^{\text{th}}$ site, $S_n$, \bnd{along the chain (a) and}{} as a function of the angle $\phi$ (b) after 100 time steps.}
\end{figure}

\begin{figure}[t]
\centering
\includegraphics[scale=0.53]{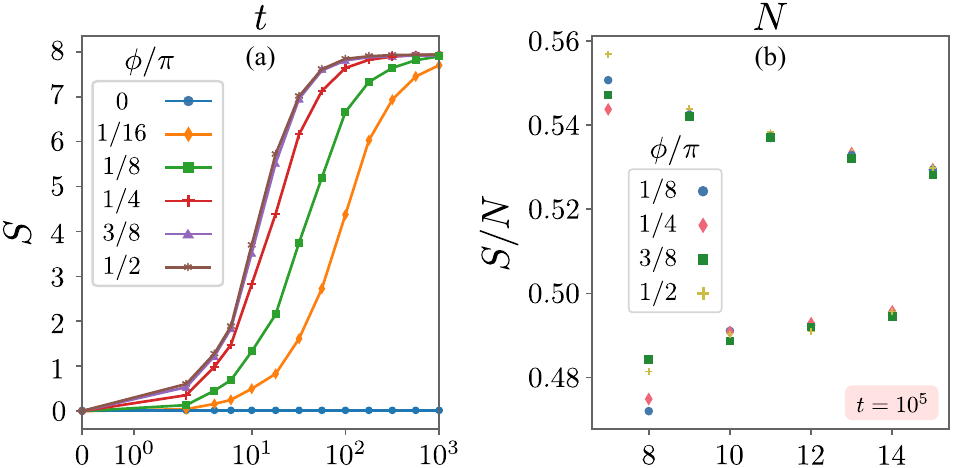}
\caption{
\label{fig:A2} 
(a) Entanglement entropy as a function of time for a lattice with $N=15$ for different values of $\phi$. (Partitioning of the system is done in half, i.e. $7+8=15$, and only spins on the left are traced out for these calculations.) For $\phi>\pi/4$, $S$ grows as $\sim \log t$ in the second decade of the evolution before the systems eventually thermalize. (b) Entanglement entropy per site $S/N$ as a function of system size $N$ after thermalization at $t=10^5$ approaches a constant indicating a volume law.}
\end{figure}

We perform a numerically exact time evolution employing the above ansatz up to $t=100$ time steps. In our simulations, we choose the truncation tolerance as \bnd{$\delta = 10^{-8}$}{$\delta = 10^{-15}$}~\footnote{\bnd{}{For example for a system of 200 sites evolved for 100 time steps, the maximum numerical error in the 2-norm of the state due to a tolerance of $10^{-15}$ is on the order of $10^{-11}$, since truncation process is repeated for each bond at each time step. We note that for the tolerance values below $10^{-7}$ our results do not change significantly. }}. The maximum bond dimension at the end of the simulations becomes $D \sim 900$ for the bond in the middle of the chain where the quantum walker starts at $t=0$. As seen from Eq.~\eqref{eq:EEmps}, bond dimension and entanglement entropy are directly related. 
\bnd{In Fig.~\ref{fig:A1}, we show how the von Neuman entropy between bi-partite partitions at the $n^{\text{th}}$ site, $S_n$, changes along the chain at $t=100$. For all the non-zero values of the interaction strength, $S_n$ is maximized around the initial site, and drops to zero near both ends (Fig.~\ref{fig:A1}(a)). As the interaction strength $\phi$ increases, spread of the entanglement is suppressed.}
{In Fig.~\ref{fig:A1new}, we show how the von Neuman entropy between partitions separated at the $n^{\text{th}}$ site, $S_n$, changes along the chain as a function of time. Since the local spins do not interact directly, the spread of the entanglement entropy is mediated by the walker. For $\phi<\pi/4$, ballistic spread of the walker causes ballistic spread of the entanglement as well, whereas partial localization of the walker yields a slower entanglement spread near the origin. As the interaction strength $\phi$ increases further, ballistic spread of the entanglement is suppressed and the entanglement entropy grows sub-linearly in time. For a quantitative analysis, we assumed the time dependence of $S(t)$ to be a combination of logarithmic and linear functions of time, so that
\begin{equation}
    S(t) = A \log{t} + B t.
    \label{eq:entanglement_vstime}
\end{equation}
In Fig.~\ref{fig:eEnt}~(a), we show how the coefficients $A$ and $B$ change with the interaction strength, $\phi$. The contribution of the logarithmic growth, $A$, increases with $\phi$, whereas the linear contribution, $B$, decreases. Therefore, as the walker gets more localized, the sub-linear growth of the entanglement prevails.} 
In Fig.~\ref{fig:eEnt}~(b), we show how $S_n$ varies with $\phi$ for different partitionings of the spin chain at $t=100$. The result for the partitioning at the origin (site $n=0$) is shown in Fig.~\ref{fig:fig4}(b), as well. When the chain is split in the middle, $S_n$ monotonically increases with increasing $\phi$. For $n\ne 0$, $S_n$ depends on the localization of the quantum walker. Near the origin, $S_n$ is generally higher in the regime where the walker is partially localized, whereas it converges to its minimum value when the walker is exponentially localized. 

We also note that for finite system calculations the entanglement entropy saturates at values that scale with the size of the system which implies a volume law for the model. The saturation of $S(t)$ is shown in Fig.~\ref{fig:A2}(a) for different values of $\phi$ and a system with $N=15$. For this system size, the saturation takes $t \sim 10^2-10^3$ depending on the localization of the walker. The strongly localized ($\phi>\pi/4$) systems saturate faster than partially localized ($\phi<\pi/4$) ones but the saturation value is determined by the system size. The scaling of the long-time value of $S$ with $N$ is shown in Fig.~\ref{fig:A2}(b) where the entropy per sites converges to a constant value.

\begin{figure}[t]
\centering
\includegraphics[scale=0.52]{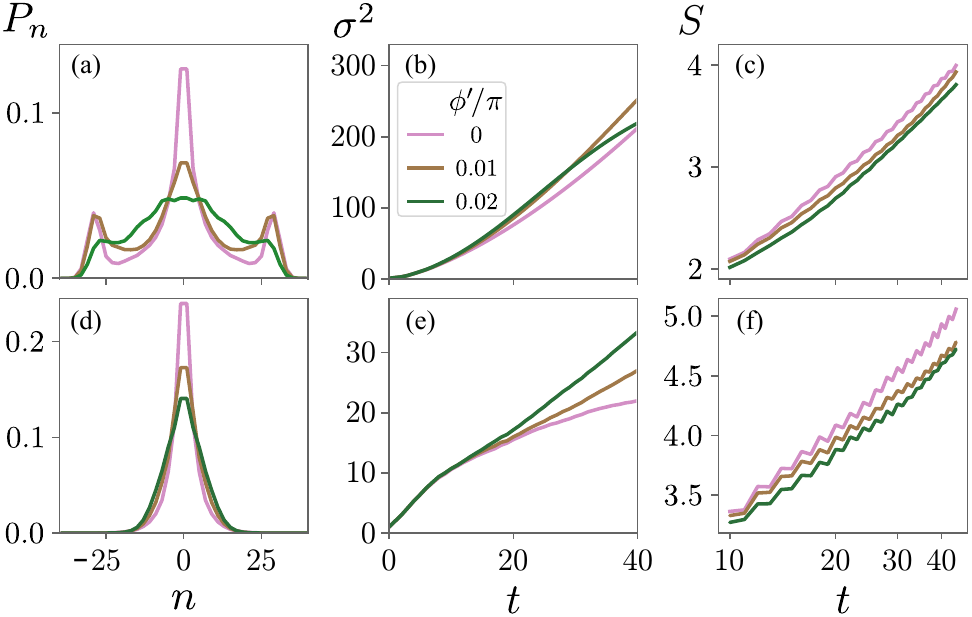}
\caption{
\label{fig:B1} Effect of the symmetry breaking field in Eq.~\eqref{eq:fieldop} for the interaction angles $\phi = \pi/8$ (upper panel) and $\phi = 3\pi/8$ (lower panel). (a) and (d) show the probability distribution. Time behavior of the variance in position space is shown in (b) and (e). Time behavior of the entanglement entropy is shown in (c) and (f).}
\end{figure}

\section{\label{app:field}Effect of a symmetry breaking field} 

Disorder-free localization (DFL) is strongly related to the extensive number of conserved quantities~\cite{smith2017disorder}. In this section, we observe the effect of a symmetry breaking field taken as a uniform field along the z-direction. We apply a field operator 
\begin{equation}
F = \prod_n e^{-i \phi' \hat{Z}_n }
\label{eq:fieldop}
\end{equation}
at every time step and consider an evolution of 40 steps from the same initial state as in the main text. We concentrate on two interaction angles. At $\phi = \pi/8$ (Fig.~\ref{fig:B1} upper panel), localization is partial and the tails at the two ends persist. When a small field such as $\phi' = \pi/100$ is present, the localization around the initial state decreases substantially, but the tails are not affected too much (Fig.~\ref{fig:B1}(a)). As we increase $\phi'$ further, the distribution becomes more uniform around the initial site and it drops around the tails, as well. Due to the diminishing of the localization with the uniform field, for the first 30 time steps variance is higher compared to the $\phi' = 0$ case and it is approximately the same for both field angles. However, for $\phi' = 2\pi/100$, the probabilities around the tails also drop yielding a slower spread and, therefore, a lower variance (Fig.~\ref{fig:B1}(b)) which is similar to a diffusive behavior. We now look at an angle for which the walker is completely localized (Fig.~\ref{fig:B1} lower panel). At $\phi = 3\pi/8$, the effect of the symmetry breaking field on the localization is lessened compared to the previous example. Though localization is diminished, it still persists for both field angles (Fig.~\ref{fig:B1}(d)). As the field strength increases, time behavior of the variance becomes similar to classical diffusion, where variance increases linearly in time (Fig.~\ref{fig:B1}(e)). For both interaction angles, the symmetry breaking field does not change the logarithmic growth of the entanglement entropy (Fig.~\ref{fig:B1}(c) and (f)).

The persistence of localization and the logarithmic growth of entanglement entropy is likely to be related to the phenomenon of prethermalization explored in integrable models~\cite{berges2004,gring2012,kollar2011,mallayya2019,mori2018}.
In these Hamiltonian models, a small perturbation is added to an exactly integrable Hamiltonian.
In this perturbed limit, the integrals of motion remain approximately conserved up to a time given by the Fermi-golden rule. In this model we expect something similar to happen, but since we are working with a unitary pertubation instead of an Hamiltonian pertubation, Fermi-golden rule does not directly apply.

To obtain the time scale at which which the perturbation starts to relax the conserved quantites $\hat{X}_n$ we expand the new perturbed dynamics around the original dynamics as follows:
\begin{eqnarray}
	U^{t}=(\hat{W}_MF)^{t} \equiv (\hat{W}_M)^{t}F^{t}+\sum_k {\phi'}^{k} \widetilde{F}_{t,k}
\end{eqnarray}
where $U = W_M F = W M F = T C M F$. Note that $M$ and $F$ can also be written as
\begin{eqnarray}
	M &=& \sum_{n=1}^N \hat{I}_2\otimes \outerp{n}{n} \otimes 
	                 ( \cos\phi\,  \hat{I}_{2^N} - i\sin\phi\,  \hat{X}_n ) \\
	F &=& \prod_{n'=1}^N \hat{I}_2\otimes \hat{I}_N \otimes 
			 ( \cos\phi'\, \hat{I}_{2^N} - i\sin\phi'\, \hat{Z}_{n'})
\end{eqnarray}
To find the first order expressions which will allow us to make a similar approximation as Fermi's golden rule, we need to be able to compute the commutator:

\onecolumngrid

\begin{eqnarray}
\left[F,M\right] = 2i\sin\phi \sum_{n}\sin\phi' \outerp{n}{n} \hat{Y}_n
     \prod_{n'\neq n}\left( \cos\phi' -i\sin\phi'\hat{Z}_{n'} \right) 
     \equiv \sin\phi'\, V
\end{eqnarray}
We then work out $\widetilde{F}_{t,k}$ recursively. 
Since $\hat{U} = \hat{W}_M F$, $\widetilde{F}_{1,k} = 0$.
Starting with $U^2$ we get:
\begin{eqnarray}
	U^2&=&\hat{W}_M^2{F}^2+\sin\phi'\, \hat{W}_M  \hat{W} V  F \\ \nonumber
\end{eqnarray}
which encourages us to define $(\sin \phi')^s \, W_{s}$ as the 
$s^\mathrm{th}$ commutator of $F$ with $\hat{W}_M$ such that 
$\sin\phi'\, W_s = [F,W_{s-1}]$ and $\sin\phi'\,W_1=[F,W_M] = W [F,M] = \sin\phi'\, W V$.
Then using:
\begin{eqnarray}
	[F,W_M^{t}]= \sin\phi'\, \sum_{k=1}^{t}  W_M^{k-1} W_1 W_M^{t-k}
\end{eqnarray}
We can get higher order terms
\begin{eqnarray}
	U^{2} &=& W_M F W_M F = W_M^2 F^2+\sin\phi'\, W_M W_1 F \nonumber \\
        U^{3} &=& W_M F U^2 
	       = W_M F \left(W_M^2 F^2 + \sin\phi' W_M W_1 F \right)  \nonumber \\
	      &=& W_M^3 F^3 + W_M [F,W_M^2] F^2 
	      + \sin\phi' W_M^2 F W_1 F 
	      + (\sin\phi')^2 W_M W_1^2 F 
	      \nonumber \\
	      &=& W_M^3 F^3 + \sin\phi' \left( 2 W_M^2 W_1 + W_M W_1 W_M \right) F^2 
	      + (\sin\phi')^2 \left( W_M^2 W_2 + W_1^2 \right) F
	      \nonumber \\
        U^{4} &=& W_M F U^3 \nonumber \\
	      &=& W_M^4 F^4 + \sin\phi' \left( 3 W_M^3 W_1 + 2 W_M^2 W_1 W_M + W_1 W_M^2 \right) F^3 
	      + (\sin\phi')^2 \left(\dots\right) F^2
	      + (\sin\phi')^3 \left(\dots\right) F
	      \nonumber 
\end{eqnarray}
Extrapolating we get the first correction as:
\begin{eqnarray}
    U^t=W_M^tF^t + \sin\phi' \sum_k^t k W_M^{k-1} W_1 W_M^{t-k} 
                 + (\sin\phi')^2 \left(\dots\right) 
\end{eqnarray}

\twocolumngrid

This term will play an import role when $\sin(\phi')\sum_{k} k \approx t^2 \sin(\phi') \approx 1$ or $t=1/\sqrt{\sin(\phi')}$. This is going as a square root because we are generating these terms both from when we commute  $[F,W^{t-1}]$ and the perturbative contribution from previous $t$.  This is true for all orders, so we generally expect those to play a role when $t\approx\sin(\phi')^{-k/2}$.

Lets now consider what each of these terms are doing. Notice that $W_1(\phi')=W_M V(\phi')$  also must be expanded in $\sin(\phi')$. The zeroth order expansion in $\sin(\phi')$ yields a term very similar to $M$ but now with $Y_n$ instead of $X_n$. At second order, $X_n$ is replaced by $Y_n+i\sum_{j\neq n}Z_j$. Working in a basis in which this operator is diagonal. The phase being imprinted onto the walker at site $n$ ends up being a sum over the expectations of $Z_j$. This will depend on how the wave function in each of the $X$ trajectories is dephasing and will likely result in the expectations of $Z_j$ being random. If this is the case, this operator averages out to $0$ and $W_1$, at second order in $\sin(\phi')$, will not contain the localizing effect $W_M$ does. Therefore it should be when the first order perturbation in $V$, (second in $U$) start to play a role that the system delocalizes. This gives
\begin{eqnarray}
    t_c \approx 1/\sin\phi'
\end{eqnarray}
This is roughly consistent with what we see in Fig.~\ref{fig:B1}, where for $\phi'=0.02\pi$, the curve departs from the $\phi'=0$ line at $t=15$ and around $t=30$ for $\phi'=0.02$. A better test requires evolving to longer times.

\bibliography{bibliography}

\end{document}